\documentclass[a4paper,11pt]{article}
\pdfoutput=1 

\usepackage{jinstpub} 

\usepackage{caption}
\usepackage{subcaption}

\usepackage{appendix}
\usepackage{upgreek}

\newcommand{\nech}{\ensuremath{\mathrm{Ne/(5\%CH_4)}}}
\newcommand{\arch}{\ensuremath{\mathrm{Ar/(5\%CH_4)}}}
\newcommand{\arco}{\ensuremath{\mathrm{Ar/(7\%CO_2)}}}

\newcommand{\lldetsize}{\ensuremath{\mathrm{50\times 50~cm^2~}}}
\newcommand{\ldetsize}{\ensuremath{\mathrm{30\times 30~cm^2~}}}
\newcommand{\sdetsize}{\ensuremath{\mathrm{10\times 10~cm^2~}}}

\title{In-beam evaluation of a medium-size Resistive-Plate WELL gaseous particle detector}

\author[a]{L. Moleri,\note{Corresponding author.}}
\author[b]{F. D. Amaro,}
\author[c]{L. Arazi,}
\author[d]{C. D. R. Azevedo,}
\author[a]{A. Breskin,}
\author[a]{A. E. C. Coimbra,}
\author[e]{E. Oliveri,}
\author[d]{F. A. Pereira,}
\author[a]{D. Shaked Renous,}
\author[a]{J. Schaarschmidt,}
\author[b]{J. M. F. dos Santos,}
\author[d]{J. F. C. A. Veloso,}
\author[a]{and S. Bressler}

\affiliation[a]{Department of Particle Physics and Astrophysics, Weizmann Institute of science,\\
7610001 Rehovot, Israel\\}
\affiliation[b]{LIBPhys, Department of Physics, University of Coimbra,\\Rua Larga, PT3004-516 Coimbra, Portugal\\}
\affiliation[c]{Physics Core Facilities, Weizmann Institute of science,\\
7610001 Rehovot, Israel\\}
\affiliation[d]{I3N, Physics Department, University of Aveiro,\\3810-193 Aveiro, Portugal\\}
\affiliation[e]{CERN,\\Meyrin, Switzerland\\}

\emailAdd{luca.moleri@weizmann.ac.il}

\abstract{In-beam evaluation of a fully-equipped medium-size \ldetsize Resistive Plate WELL (RPWELL) detector is presented. It consists here of a single element gas-avalanche multiplier with Semitron ESD225 resistive plate, 1~cm$^2$ readout pads and APV25/SRS electronics. Similarly to previous results with small detector prototypes, stable operation at high detection efficiency (>98$\%$) and low average pad multiplicity ($\sim$1.2) were recorded with 150~GeV muon and high-rate pion beams, in \nech, \arch~and \arco. This is an important step towards the realization of robust detectors suitable for applications requiring large-area coverage; among them Digital Hadron Calorimetry.}


\begin{document}
\maketitle
\flushbottom

\section{Introduction}
\label{sec: Introduction}

The Resistive Plate WELL (RPWELL)~\cite{rubin2013first} is a single-faced (Copper-clad on one side) Thick Gas Electron Multiplier (THGEM)~\cite{chechik2004thick,breskin2009concise}, coupled to a segmented readout anode through a high bulk resistivity plate (figure~\ref{fig: RPWELL scheme}). Extensive laboratory studies of small RPWELL detector prototypes operated in \nech~\cite{rubin2013first} have been performed; they were followed by larger (\sdetsize) detector investigations, with pad readout, in muon and pion beams. These were operated in \nech~\cite{bressler2016first}, and in \arch~and \arco~\cite{moleri2016resistive}. These and other studies of THGEM-like detectors (summarized in~\cite{bressler2013recent}), aimed at validating the potential applicability of the RPWELL as sampling element in digital hadron calorimetry (DHCAL). They have demonstrated discharge-free operation at high gas-avalanche gains, over a broad dynamic range. The results summarized in~\cite{bressler2016first,moleri2016resistive}, were obtained with detectors having a 5~mm conversion/drift gap, followed by a single thin (0.8~mm) multiplier, coupled through a 0.4~mm thick Semitron$^{\circledR}$ ESD225\footnote{www.quadrantplastics.com} resistive plate - to an anode segmented into 1~cm$^2$ pads; charge signals from the pads were recorded by a single APV25 chip~\cite{french2001design} and SRS readout system~\cite{martoiu2013development}. Detection efficiency values greater than 98$\%$ were reached, at low average pad multiplicity values of $\sim$1.2 - in all three gas mixtures; moreover, in these conditions, the RPWELL detector displayed no discharges, also under a high pion flux. Constant detection efficiency was recorded up to a pion flux of 10$^4$~Hz/cm$^2$, decreasing by a few percent at $\sim$10$^5$~Hz/cm$^2$. These former results, obtained with small detector prototypes, suggested that RPWELL detectors are promising for applications that require cost-effective solutions for large-area coverage. For example, the Digital, or Semi-Digital Hadron Calorimeter ((S)DHCAL)~\cite{adams2016design,elmahroug2014study}, foreseen for the SiD experiment in the future international linear collider (ILC)~\cite{behnke2013international}, the overall instrumented area will be as large as $\sim$4000~m$^2$. Another application could be photon detection for large-area UV-RICH detectors~\cite{chechik2005thick,alexeev2013thgem}. Previous studies with large area THGEM detectors are described in~\cite{xie2013development} and in~\cite{alexeev2016status}; in the latter, two \ldetsize~double-faced THGEM electrodes in cascade are followed by a MICROMEGAS. Although this configuration is different than the RPWELL one, common aspects were considered. These include the mechanical support used in the detector assembly and various quality criteria imposed on the THGEM electrodes themselves. 
Motivated by these results, in this work we address the challenge of scaling up  the detector dimensions. For the first time, a fully equipped medium-size (\ldetsize) RPWELL detector was assembled and investigated in muon and high-rate pion beams, with Neon- and Argon-based gas mixtures.
The experimental setup and methodology are described in section~\ref{sec: Experimental setup and methodology}. The results are presented in section~\ref{sec: Results} followed by a summary and discussion in section~\ref{sec: Summary and discussion}. 

\section{Experimental setup and methodology}
\label{sec: Experimental setup and methodology}
\subsection{The 30$\times$30 cm$^2$ RPWELL detector prototype}
\label{sec: prototype}
\begin{figure}
\centering
\includegraphics[scale=0.3]{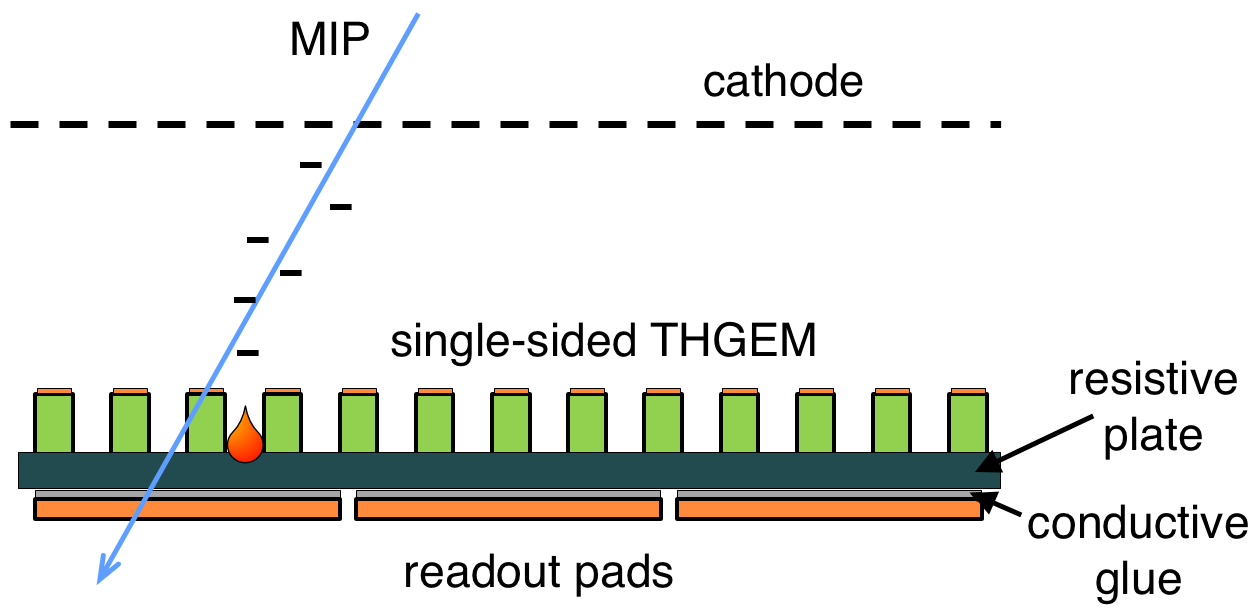}
\caption{The RPWELL operation principle.}\label{fig: RPWELL scheme}
\end{figure}

\begin{figure}[h]
\centering
\begin{subfigure}[t]{0.33\linewidth}\caption{Single-sided THGEM}
\includegraphics[scale=0.023]{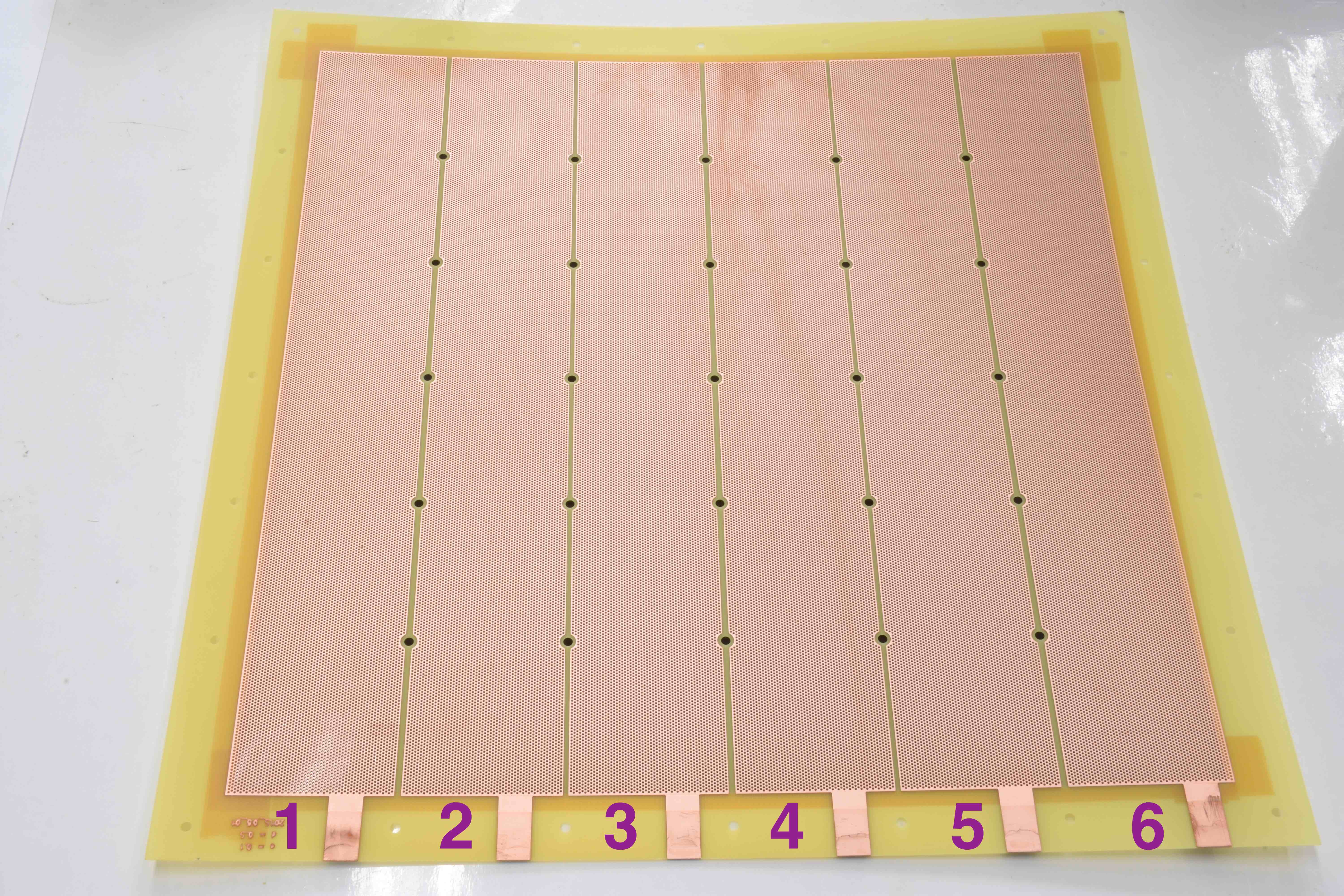}
\end{subfigure}
\begin{subfigure}[t]{0.3\linewidth}\caption{Readout anode}
\includegraphics[scale=0.05]{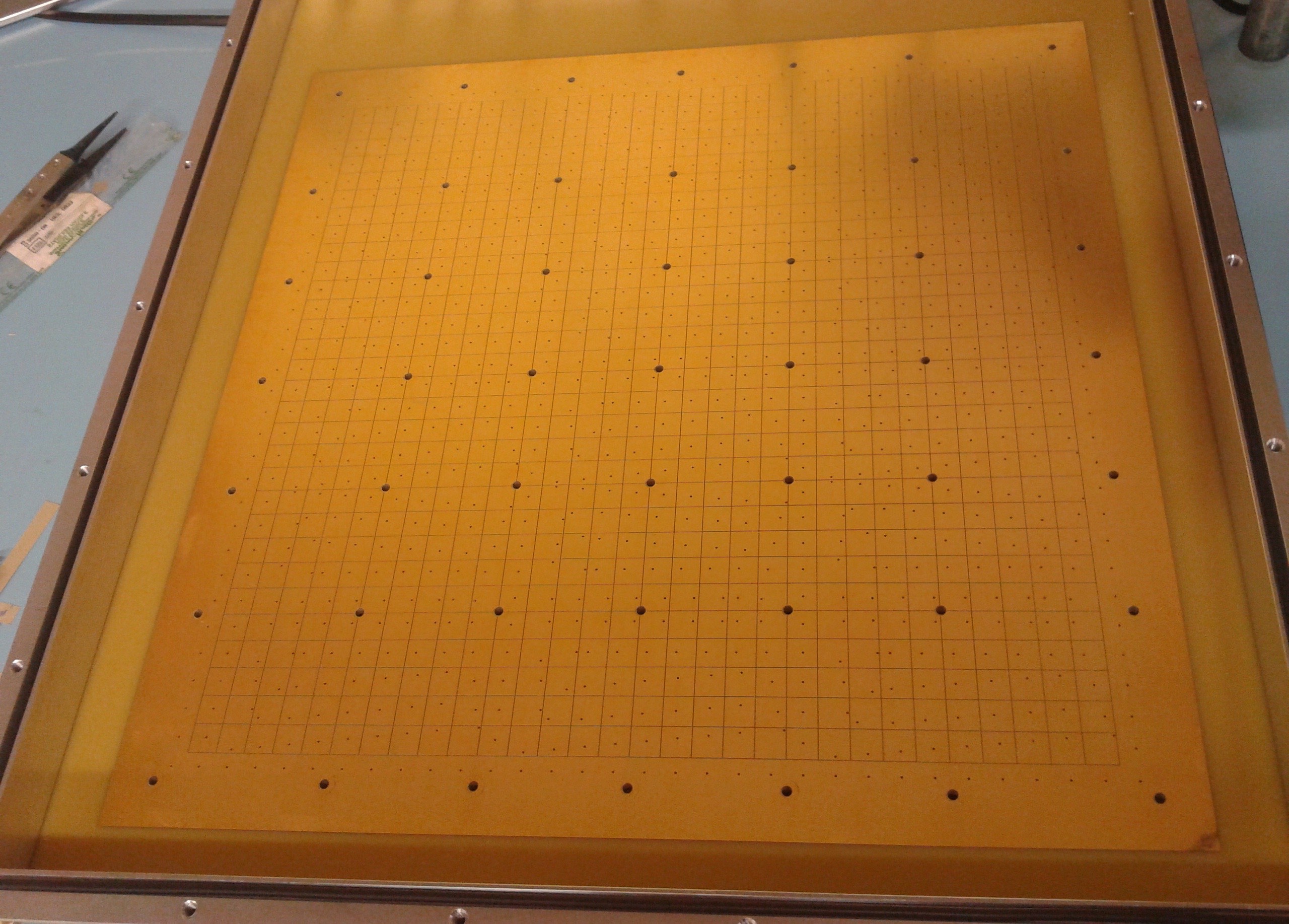}
\end{subfigure}
\begin{subfigure}[t]{0.33\linewidth}\caption{Resistive plate}
\includegraphics[scale= 0.05]{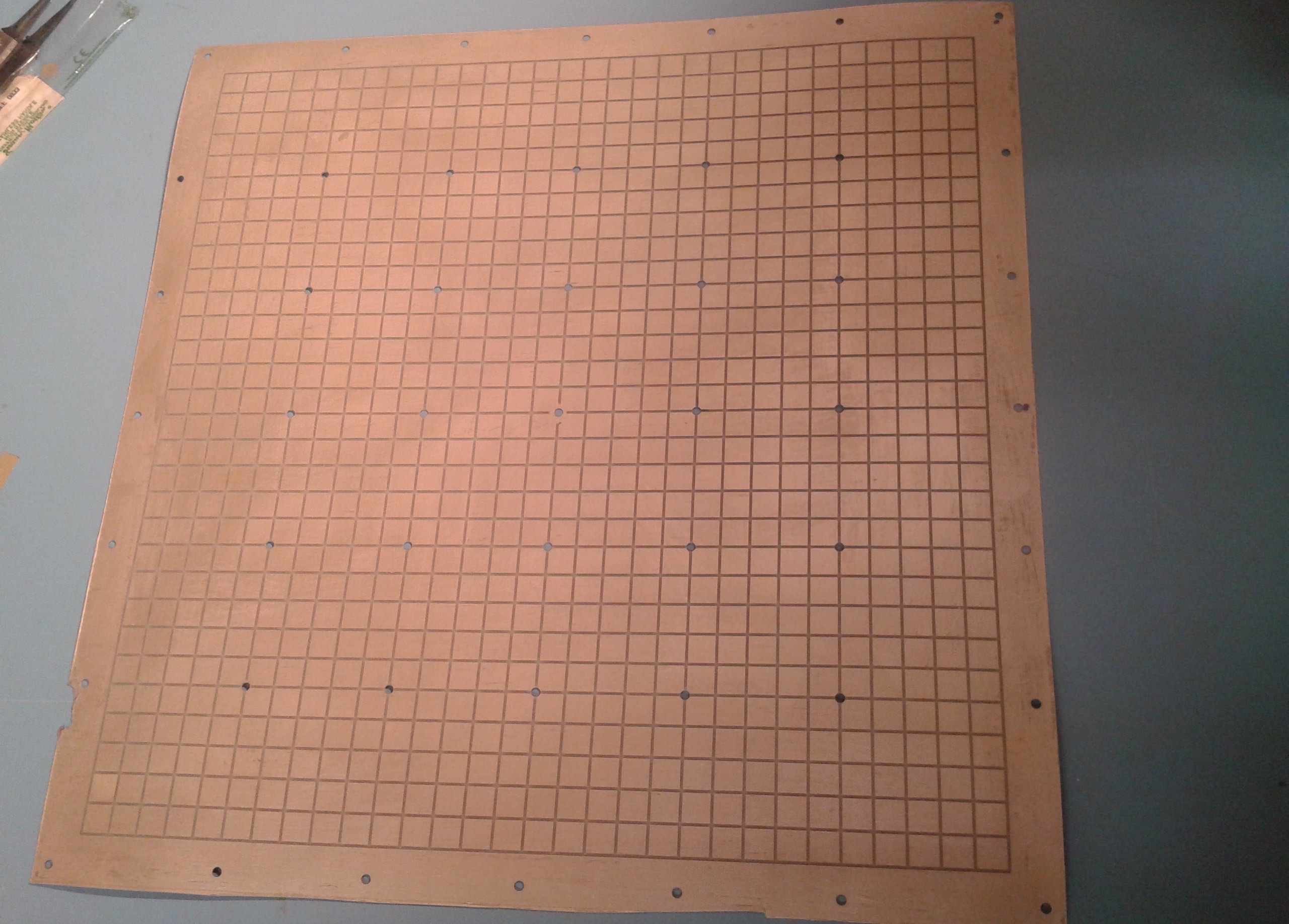}
\end{subfigure}\vspace{5mm}
\begin{subfigure}[t]{0.4\linewidth}\caption{}
\includegraphics[scale= 0.06]{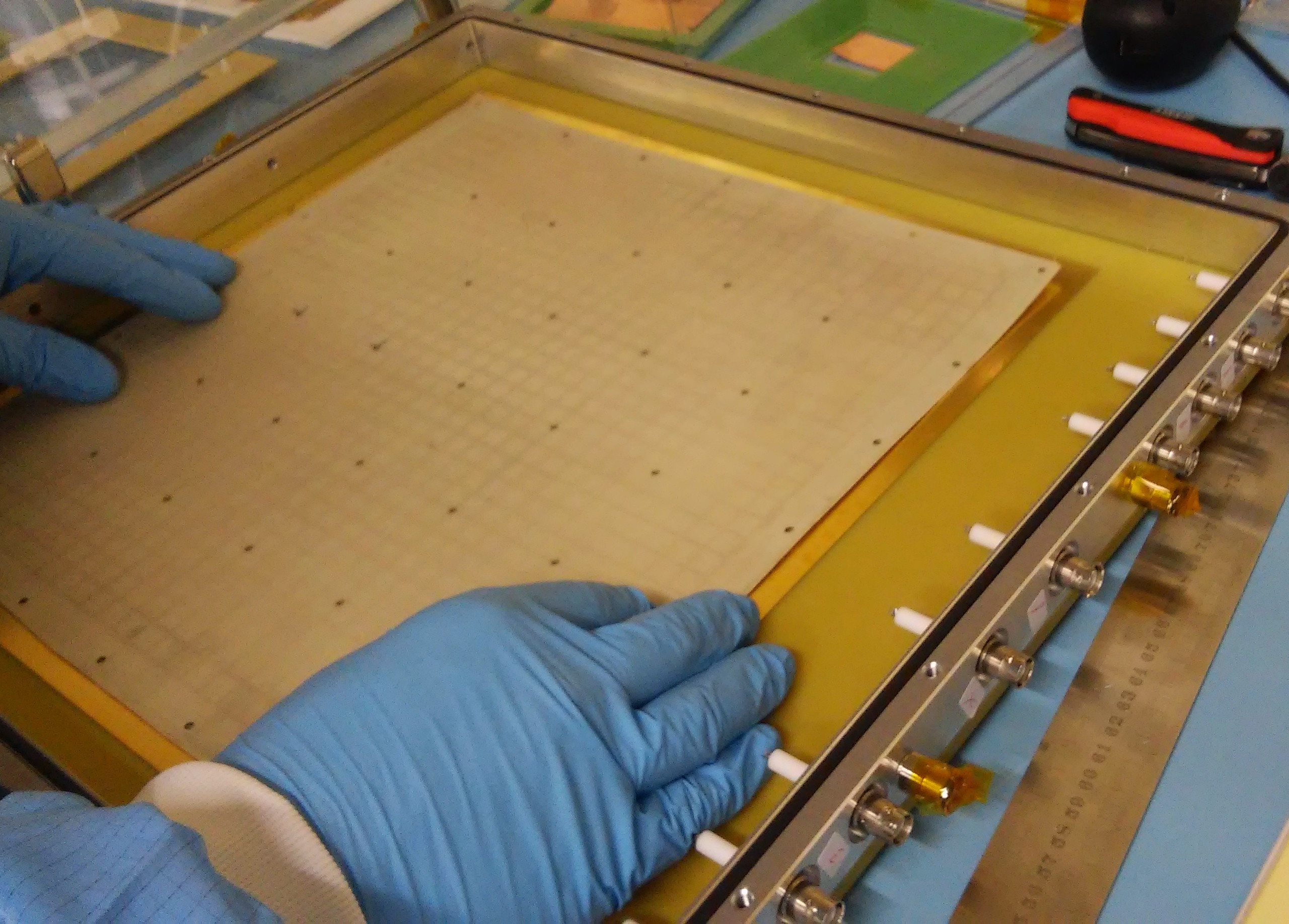}
\end{subfigure}
\begin{subfigure}[t]{0.4\linewidth}\caption{}
\includegraphics[scale=0.05]{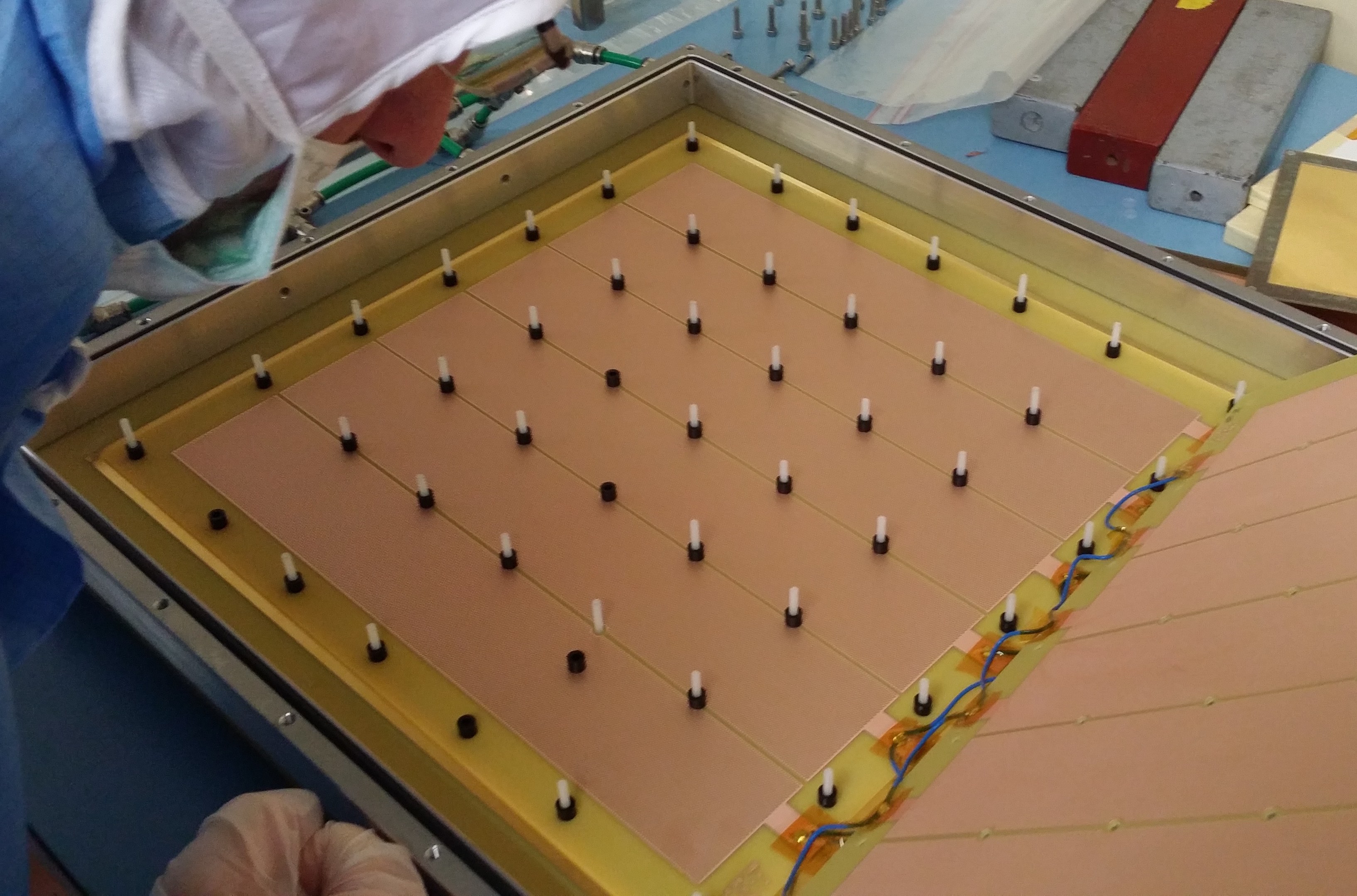}
\end{subfigure}
\caption{Detector prototype parts: (a)-(c). (d) Assembling the resistive plate (c) on top of the readout anode (b), using conductive tape. (e) The open detector with all its elements (except the vessel cover): the anode and resistive plate (not visible); the THGEM electrode, with the support nylon pins (white) and Delrin$^{\circledR}$ spacers (black); the cathode (lifted on the right side); the aluminium vessel.}\label{fig: assembly}
\end{figure}

The RPWELL simplified scheme is shown in figure~\ref{fig: RPWELL scheme}; a single-sided THGEM electrode is coupled to the readout anode through a resistive plate, preceded by a conversion/drift gap and a cathode. Various detector parts and their assembly are shown in figure~\ref{fig: assembly}.

\begin{figure}
\begin{minipage}[t]{0.4\linewidth}
\begin{subfigure}[t]{0.5\linewidth}\caption{}
\includegraphics[scale=0.45]{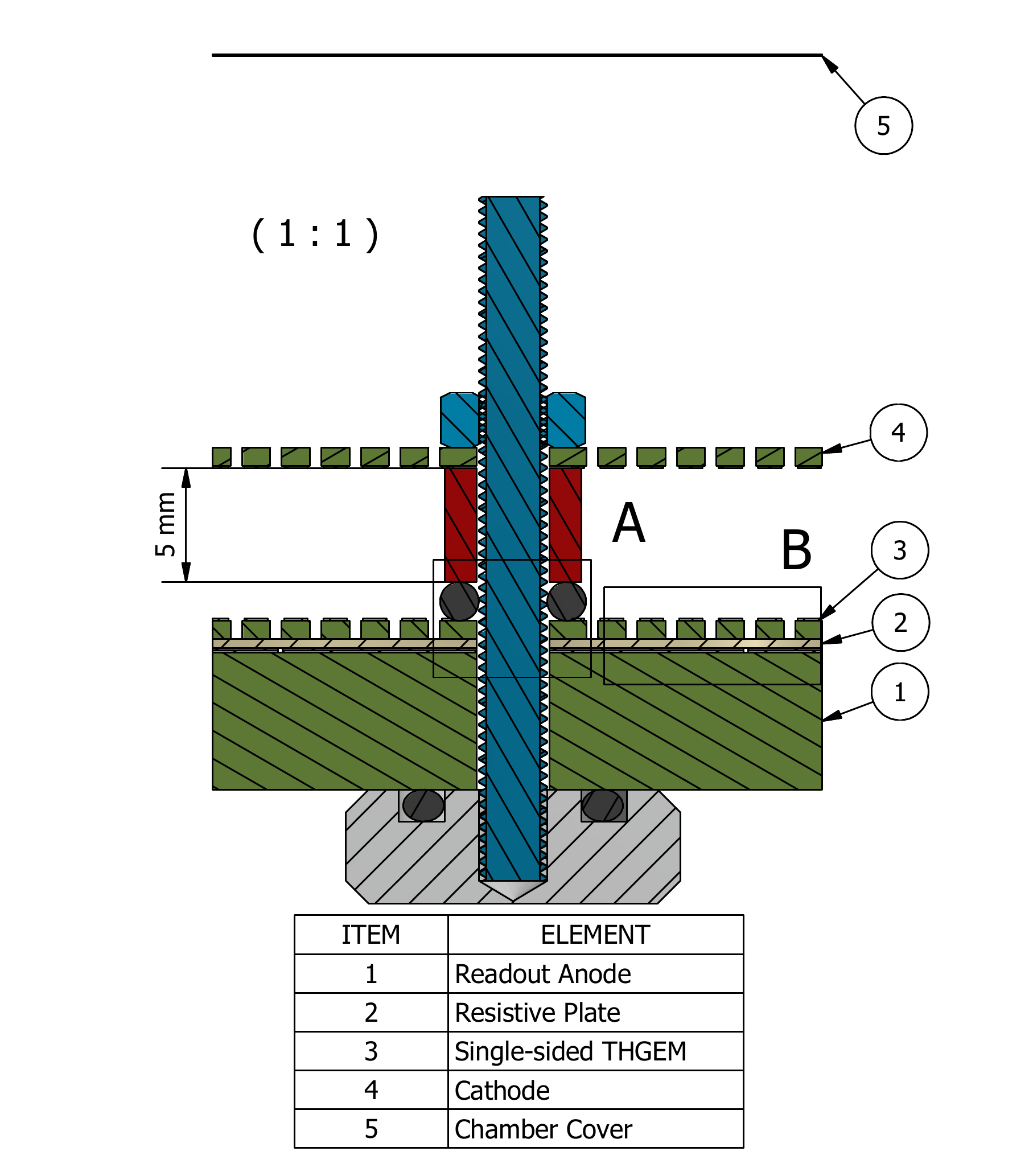}
\end{subfigure}
\end{minipage}
\begin{minipage}[t]{0.5\linewidth}
\centering
\begin{subfigure}[t]
{0.5\linewidth}\caption{}
\vspace{-1cm}
\includegraphics[scale=0.45]{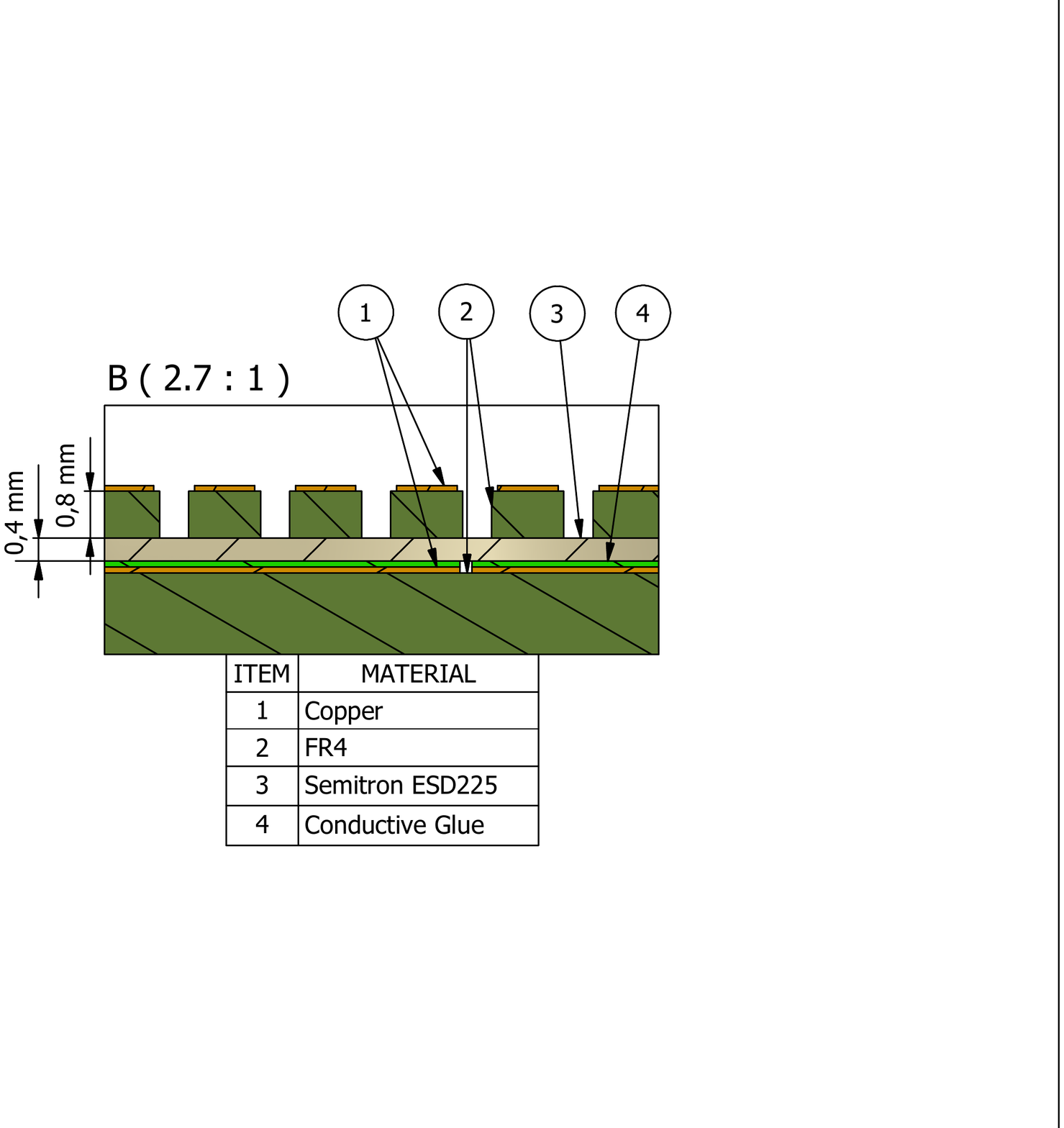}
\end{subfigure}
\begin{subfigure}[t]{0.5\linewidth}\caption{}
\vspace{-1cm}
\hspace{3cm}
\includegraphics[scale=0.45]{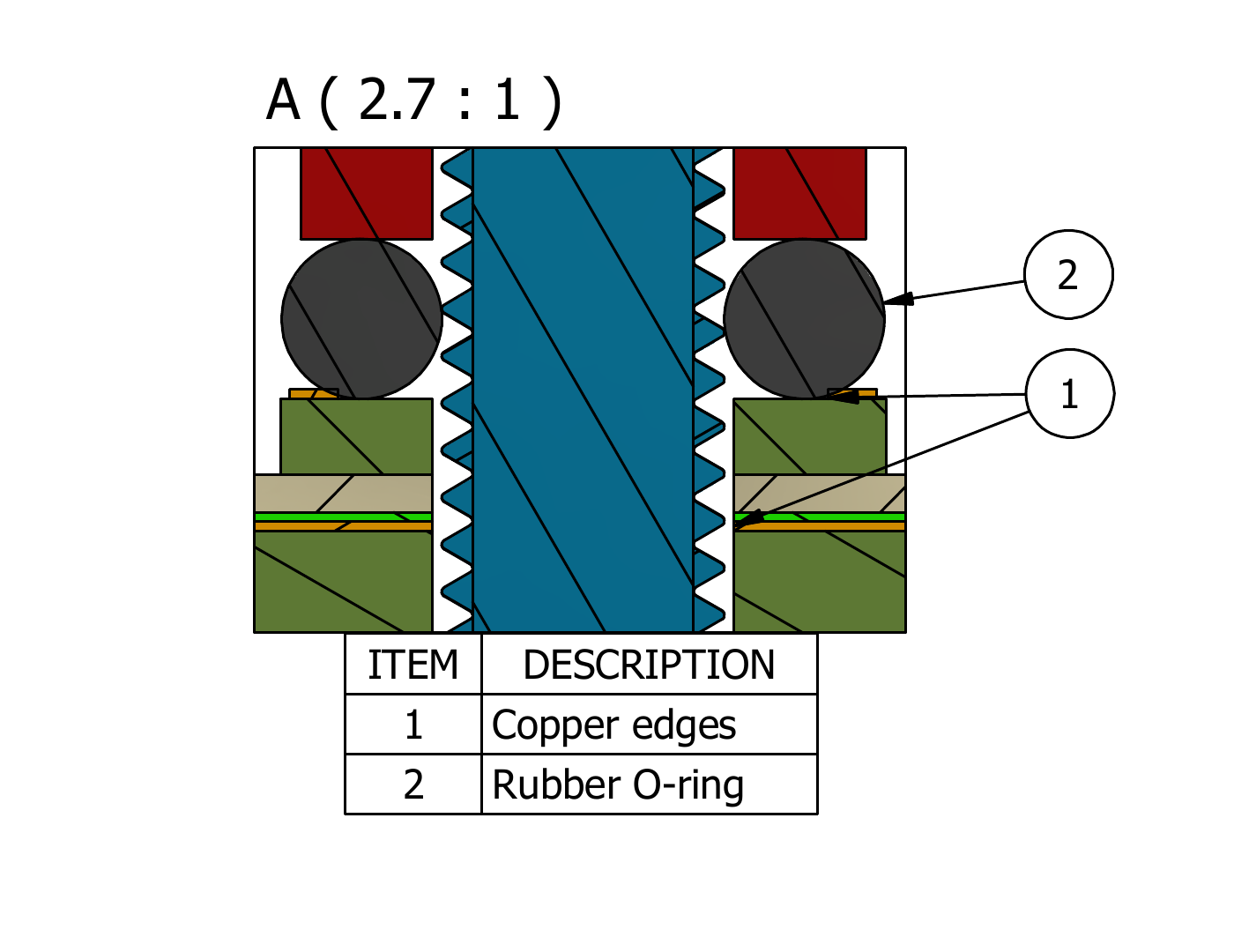}
\end{subfigure}
\end{minipage}
\caption{Mechanical design of the detector prototype assembly. (a) Section around a support pin. (b) Detail of the RPWELL multiplier, including the single-sided THGEM electrode, the resistive plate and the readout anode. (c) Zoom-in on the o-ring pressing on the THGEM electrode to close the open path between the segment edge and the anode.}\label{fig: detector scheme}
\end{figure}

Based on our previous studies~\cite{shalem2006advances,bressler2016first}, we used a single-sided Copper-clad FR4 THGEM electrode\footnote{Produced by Eltos S.P.A. (www.eltos.com), and cleaned at CERN PCB workshop} with a nominal thickness of 0.8~mm; its measured thickness (including both Copper and FR4) was 0.96~mm, with variations smaller than 40~$\upmu$m across the surface. Variations of this level ($\sim$4$\%$) were also observed in other works employing large-area THGEM electrodes~\cite{alexeev2014progresses}. The resulting gain variations (as high as 50$\%$ for the detector studied in~\cite{alexeev2014progresses}), can be mitigated in future detectors by proper selection of the FR4 sheets. The electrode had 0.5~mm diameter holes, drilled on a 1~mm pitch hexagonal pattern; chemically etched 0.1~mm rims around the holes prevented sharp edges and other eventual defects.
The \ldetsize THGEM electrode comprised six electrically decoupled 5$\times$30~cm$^2$ segments (figure~\ref{fig: assembly}-a); 3~mm gaps were left between neighboring segments, to avoid inter-segment discharges in case of significant potential drop on one of them. In this study, the inter-segment dead-area was not optimized for efficiency losses. The readout anode was composed of a 30$\times$30 matrix of 1$\times$1 cm$^2$ readout pads (figure~\ref{fig: assembly}-b); the individual pads were electrically connected to a 0.4~mm thick Semitron$^{\circledR}$ ESD225\footnote{www.quadrantplastics.com} static dissipative polymer plate of $\sim$10$^9$~$\Omega$cm bulk resistivity (figure~\ref{fig: assembly}-c). To assure good electrical contact, the bottom of the resistive material was mechanically patterned (by 1~mm wide, 50~$\upmu$m deep machined groves) into 1~cm$^2$ pads (corresponding to the metal pads of the readout electrode); each of them was coated with Pelco$^{\circledR}$ conductive silver paint\footnote{https://www.tedpella.com}. The resistive-plate pads were individually connected to corresponding readout pads with small pieces of 3M\textsuperscript{TM} Electrically Conductive Adhesive Transfer Tape 9707\footnote{www.3m.com}. Figure~\ref{fig: assembly}-d shows the positioning of the resistive plate on top of the readout anode. For practical reasons, the cathode was also a THGEM electrode of similar geometry, with all the segments interconnected. In the present work, the detector was designed in a modular way, mounted within an aluminum vessel, to permit modifications. Therefore, to assure conversion/drift gap homogeneity and good contact between electrodes, rather than using glued spacers, the detector components were assembled on an array of 49 nylon pins of 3~mm diameter (figure~\ref{fig: assembly}-e); these were fixed to the 6~mm thick padded-anode board using buttons and o-rings (figure~\ref{fig: assembly}). The pins were arranged over the active area in a square lattice with 5~cm pitch. The conversion/drift gap was determined by 5~mm diameter, 5~mm thick Delrin$^{\circledR}$ spacers inserted on the nylon pins; they were pressing on rubber o-rings, mounted underneath, against the THGEM electrode in an attempt to avoid open paths along the pins between the THGEM segment edge and the anode (see figure~\ref{fig: detector scheme}-c). This is shown to scale in the mechanical design in figure~\ref{fig: detector scheme}, and discussed in detail in section~\ref{sec: rate scan}. The cathode was placed on top of the spacers; the whole detector stack was closed with nylon nuts to ensure uniform contact between the THGEM electrode and the resistive plate.
The detector was mounted in a gas-tight vessel with gas and high-voltage (HV) feed-through connectors. The readout anode board served also as the bottom part of the vessel. The anode was grounded through the readout, while the THGEM electrode and the cathode were biased by individual channels of CAEN A1833P and A1821N HV power-supply boards, remotely controlled with a CAEN SY2527 unit. Each couple of THGEM segments was independently biased (segments 1-2, 3-4, 5-6 in figure~\ref{fig: assembly}-a), using different HV supply channels. The voltage and current in each channel were monitored and stored (20~nA resolution). All HV inputs were connected through low-pass filters. The RPWELL potential ($\Delta$V$_{RPWELL}$) with respect to the anode was varied throughout the experiment, while the drift potential was kept constant $\Delta$V$_{drift}$= 250~V, corresponding to a drift field of $\sim$0.5~kV/cm across the $\sim$5~mm conversion/drift gap. This value was chosen based on previous works~\cite{cortesi2009thgem, shalem2006advances}.
The detector was installed at the CERN-SPS H2 test beam area and investigated with $\sim$150~GeV muons and pions; it has been operated in three different gas mixtures (section~\ref{sec: gas mixtures}) at atmospheric pressure and room temperature, at a gas flow of 50-100 cc/min. No significant gain variation was observed at these different flow values.

\subsection{Gas mixtures}
\label{sec: gas mixtures}
A set of reference measurements were conducted with our previously employed~\cite{bressler2016first,moleri2016resistive} "standard" Neon mixture of \nech~prior to those with Argon-based gas mixtures: \arch~and \arco. The operation in Argon mixtures required higher electric fields - and therefore higher voltages - with respect to Neon, to reach similar gains. However, Argon mixtures present two main advantages: (1) larger average number of electron-ion pairs produced by Minimum Ionizing Particles (MIPs); e.g in 1~cm of gas in standard conditions the numbers are 94 in Argon, and 39 in Neon~\cite{sauli2014gaseous}, allowing to use a smaller conversion/drift gap maintaining high detection efficiency. (2) Argon is considerably cheaper than Neon, hence more attractive for applications requiring large-area coverage. The use of the non-flammable CO$_2$ instead  of CH$_4$ as a photon quencher could have some additional advantage.

\subsection{Tracking, readout system and analysis framework}
\label{sec: Setup}

The experiments were carried out at the CERN-RD51 beam line. The trigger and tracking system (based on the CERN-RD51 telescope~\cite{karakostas2010micromegas}), the data acquisition system (based on the SRS/APV25 readout electronics~\cite{martoiu2013development,french2001design}) and the analysis framework were the same as in~\cite{bressler2016first,moleri2016resistive}; they are described in detail in~\cite{bressler2016first}. The RPWELL chamber was placed along the beam line in between two tracker elements. The global detection efficiency was defined as the fraction of tracks matched to a pads cluster found not more than W~[mm] away from the track trajectory in both x and y directions. The average pad multiplicity was defined as the average number of pads in a matched cluster. Only pads with charge above threshold were considered. The threshold for each pad was relative to the channel noise and it was set using a common Zero-order Suppression Factor (ZSF) (for details see~\cite{bressler2016first}).
The detector's discharge probability was defined as the number of discharges divided by the number of hits in the active region of the detector (i.e., in the total area covered by the crossing beam). The number of discharges was extracted directly from the power supply log files by counting the resulting spikes in the supplied current monitor. Due to the low rate of the muon beam, only pion runs were used to estimate the discharge probability. Since pions are prone to induce highly-ionizing secondary events, this study yielded an upper limit of MIP-induced discharge probability.

The detector working point was adjusted to optimize its performance, targeting high detection efficiency at low pad multiplicity. The latter is a requirement for particle counting, e.g. in a potential application of the RPWELL as a sampling element in DHCAL~\cite{arazi2012thgem}. The optimization was done using a set of measurements with $\sim$10$^2$~Hz/cm$^2$ broad (5$\times$5~cm$^2$) muon beam and a $\sim$10$^4$~Hz/cm$^2$ narrow (2$\times$2~cm$^2$) pion beam. In both cases, only tracks hitting the detector in the central 4$\times$3~cm$^2$ beam area were considered. To fix the values of ZSF and W we followed the method described in~\cite{bressler2016first}. The optimized working points in each of the gas mixtures are summarized in table~\ref{tab: optimization}. 

\begin{table}
\centering\caption{Optimized parameters for Neon and Argon mixtures.}\label{tab: optimization}
\begin{tabular}{| l | l | l | l |}
\hline
gas & ZSF & W~[mm]\\
\hline
\nech & 15  & 15 \\
\arch & 15 & 10 \\
\arco & 15 & 15 \\
\hline
\end{tabular}
\end{table}

\section{Results}
\label{sec: Results}

\subsection{Global and local detection efficiency and average pad multiplicity}
\label{sec: Detection efficiency and pad multiplicity}

\begin{figure}
\begin{subfigure}[t]{0.45\linewidth}\caption{}
\includegraphics[scale=0.35]{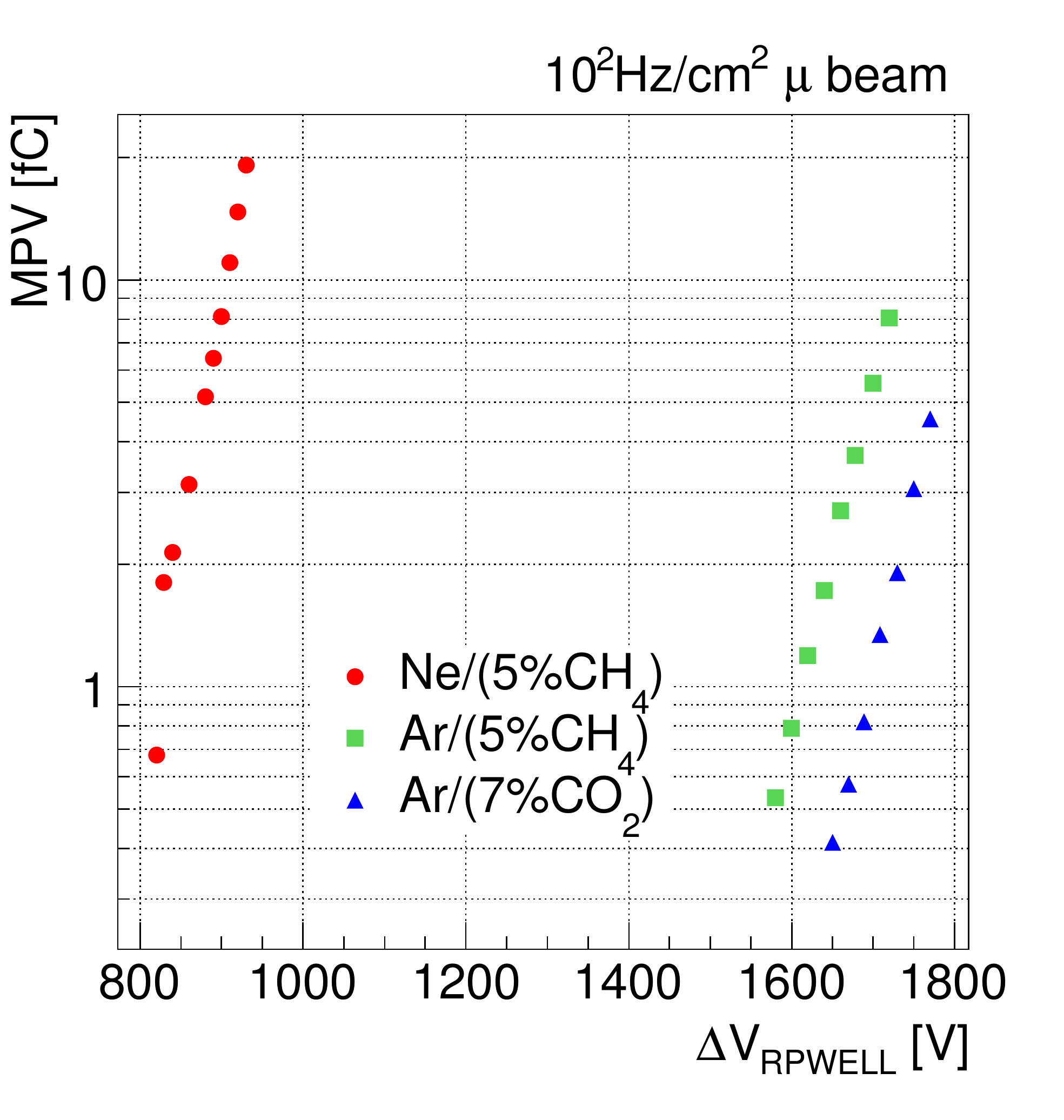}
\end{subfigure}\hspace{10mm}
\begin{subfigure}[t]{0.45\linewidth}\caption{}
\includegraphics[scale=0.35]{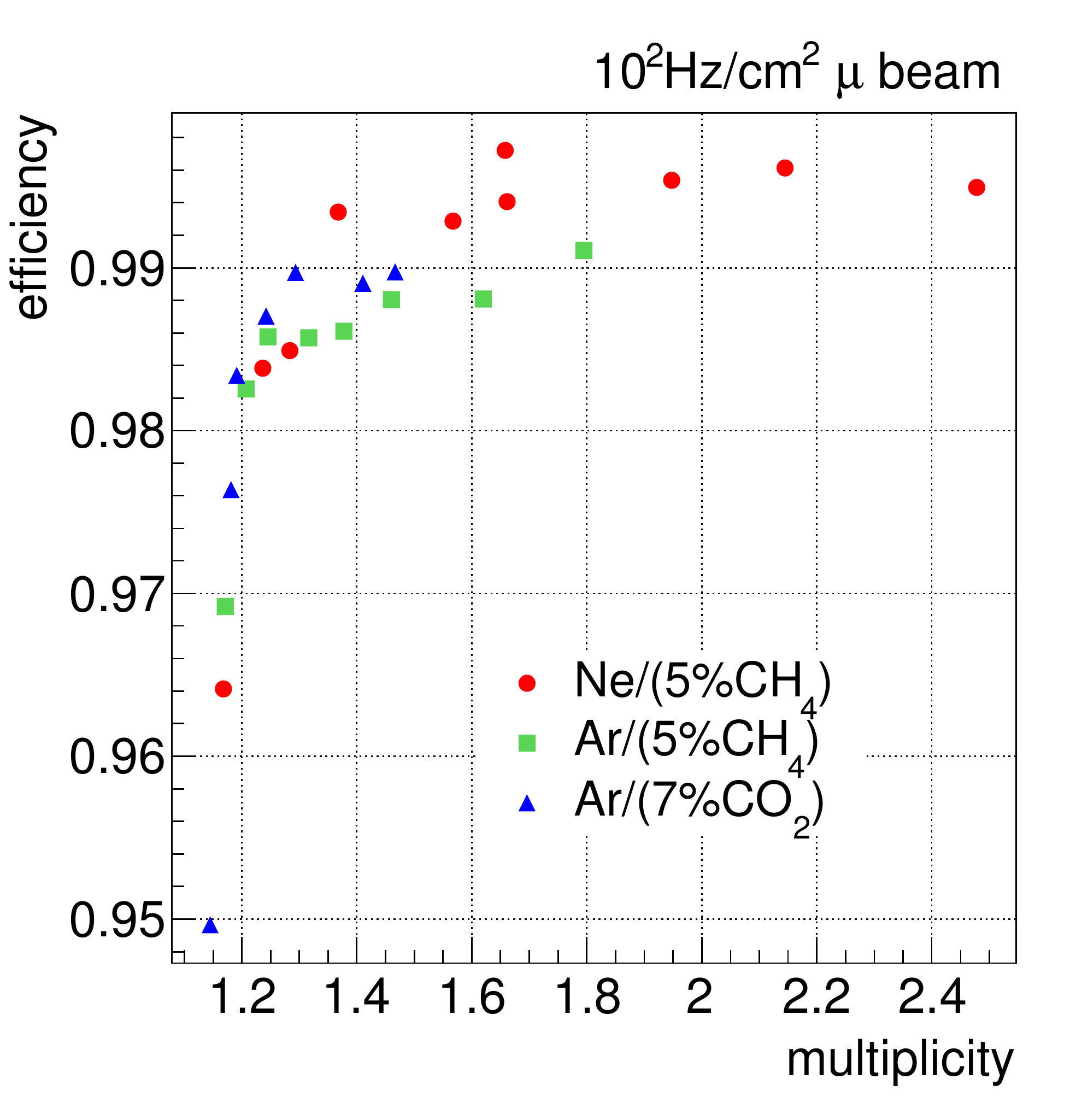}
\end{subfigure}
\caption{For the same data set: the most probable value (MPV) of the Landau spectrum as a function of $\Delta$V$_{RPWELL}$ (a) and global detection efficiency as a function of the average pad multiplicity for different $\Delta$V$_{RPWELL}$ values (b). \ldetsize RPWELL detector, operated in $\sim$10$^2$~Hz muon beam in \nech, \arch~ and \arco.}\label{fig: HV scan}
\end{figure}

\begin{figure}
\begin{subfigure}[t]{0.5\linewidth}\caption{}
\includegraphics[scale=0.35]{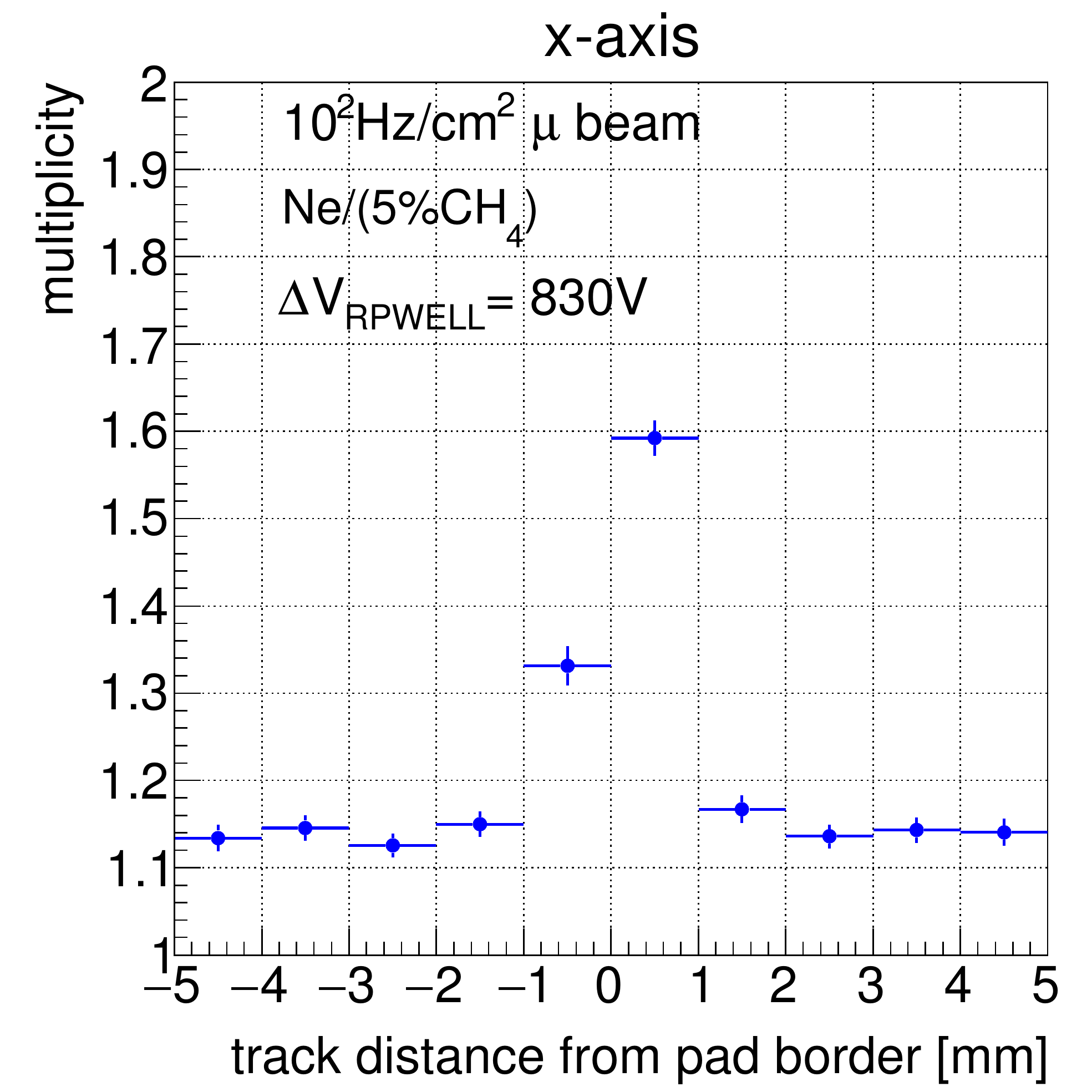}
\end{subfigure}
\begin{subfigure}[t]{0.5\linewidth}\caption{}
\includegraphics[scale=0.35]{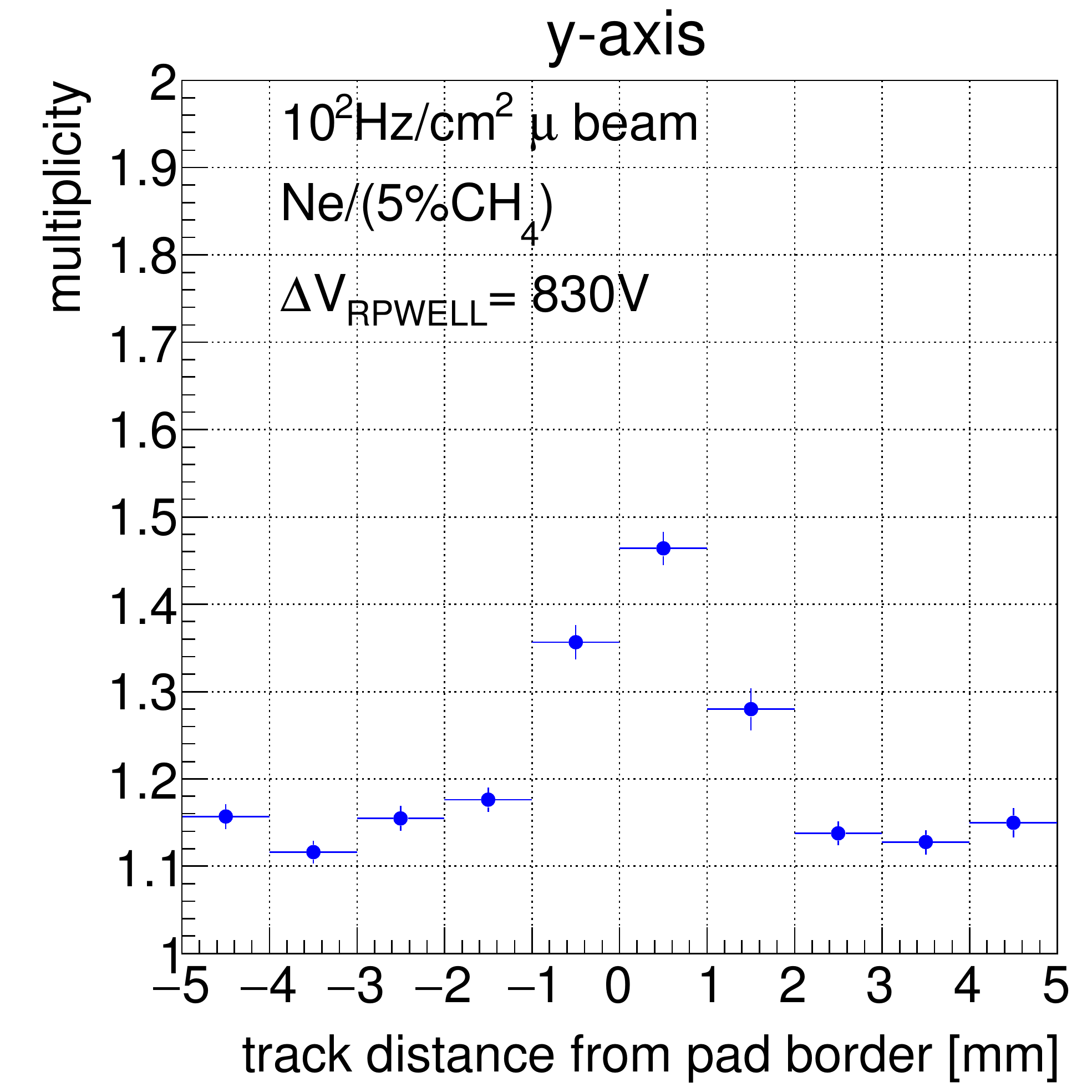}
\end{subfigure}\caption{For the same data set: the local average pad multiplicity as a function of the track distance from the pad boundary along the detector's x-axis (a) and y-axis (b) measured in the \ldetsize RPWELL detector, in \nech, with a $\sim$10$^2$~Hz/cm$^2$ muon beam. Similar results were obtained with the Argon gas mixtures.}\label{fig: local multiplicity}
\end{figure}

The detector was operated at absolute gains of the order of 10$^4$ in the three gas mixtures. As explained in~\cite{bressler2016first}, the effective gain was of the order of 10$^3$. This is due to the 75~ns shaping time of the APV25 chip, and the $\sim$1$\mu$s rise-time of the RPWELL signal, which results in integrating over $\sim$20$\%$ of the total charge. The average numbers of MIP-induced electrons deposited in the $\sim$5~mm conversion/drift gap in the Neon and Argon mixtures, were 20 and 47 respectively~\cite{sauli2014gaseous}. The most probable values (MPVs) of the Landau spectra measured for the three mixtures in a $\sim$10$^2$~Hz/cm$^2$ muon beam, are shown in figure~\ref{fig: HV scan}-a as a function of  $\Delta$V$_{RPWELL}$.
For the same data set, figure~\ref{fig: HV scan}-b depicts the global detection efficiency values as a function of the average pad multiplicity for different $\Delta$V$_{RPWELL}$ values. In table~\ref{tab: HV scan results} we summarize the optimal operation voltages and the corresponding values of global efficiency and average pad multiplicity. High detection efficiency values (98$\%$) at low pad multiplicity ($\sim$1.2) were reached in all the gas mixtures investigated.
Comparing these results to that previously obtained with the smaller RPWELL detector~\cite{moleri2016resistive}, the \ldetsize one reached optimal operation at lower potentials, due to a slightly higher gain. This could be attributed to differences in the gas pressure and circulation during the experiment, and possibly to small differences (within production tolerances) in the detector geometry.  Another observed difference is that for the same value of the global detection efficiency the larger detector had higher average pad multiplicity than the \sdetsize one. This was explained by the following observations.
In figure~\ref{fig: local multiplicity} we show the local average pad multiplicity values as a function of the track distance from the pad boundary along the x and y-axis, measured in $\sim$10$^2$Hz/cm$^2$ muon beam in \nech; similar results were obtained in the Argon mixtures. As expected, the local pad multiplicity is uniform and low ($\sim$1.1), except for a narrow region around the pad border; there, the induced signal is shared between the two neighboring pads, resulting in local pad multiplicity closer to 2. It appears that the local pad multiplicity distribution close to the pad boundary is somewhat narrower along the x-axis (figure~\ref{fig: local multiplicity}-a) compared to that along the y-axis (figure~\ref{fig: local multiplicity}-b). This difference is attributed to the hexagonal pattern of THGEM holes. In this geometry, the pad borders along the x-axis are always in front of the middle of a raw of holes, minimizing physical charge spreading; along the y-axis, the pad boundaries are located at different distances from the holes centers, causing sometimes the charge of a given avalanche to spread among different holes, belonging to different readout pads. A comparison of the results presented in figure~\ref{fig: local multiplicity} to similar ones presented in~\cite{bressler2016first}, shows a narrower increase in the local multiplicity at the pad boundary respect to the one of the smaller detector. This is attributed to the different pattern of the THGEM holes in the previous work. In~\cite{bressler2016first}, the holes  were arranged in a square pattern with a pitch of 0.96~mm; neighboring 10$\times$10 hole areas were separated by 0.68~mm plain Copper strips, at the boundaries between two readout pads. A track traversing the detector in this inter-pad region is more likely to induce charge in holes belonging to two different pads. For this reason the local multiplicity is relatively high farther away from the pad border. We emphasize that these results were obtained with tracks perpendicular to the detector plane. They demonstrate the role of the THGEM geometry in the detector performance. Further optimization is needed for improving the performance, also based on the targeted application.

\begin{table}[h]
\centering\caption{Performance of the detector at optimal parameters values.}\label{tab: HV scan results}
\begin{tabular}{| l | l | l | l |}
\hline
gas & $\Delta$V$_{RPWELL}~[V]$ & global efficiency & average multiplicity\\
\hline
\nech & 830 & 98$\%$ & 1.24 \\
 \hline
\arch & 1600 & 98$\%$ & 1.21\\
 \hline
\arco & 1690 & 98$\%$ & 1.19\\
\hline
\end{tabular}
\end{table}

\subsection{Detector performance and stability at high rates}
\label{sec: rate scan}

\begin{figure}
\begin{subfigure}[t]{0.45\linewidth}\caption{}
\includegraphics[scale=0.35]{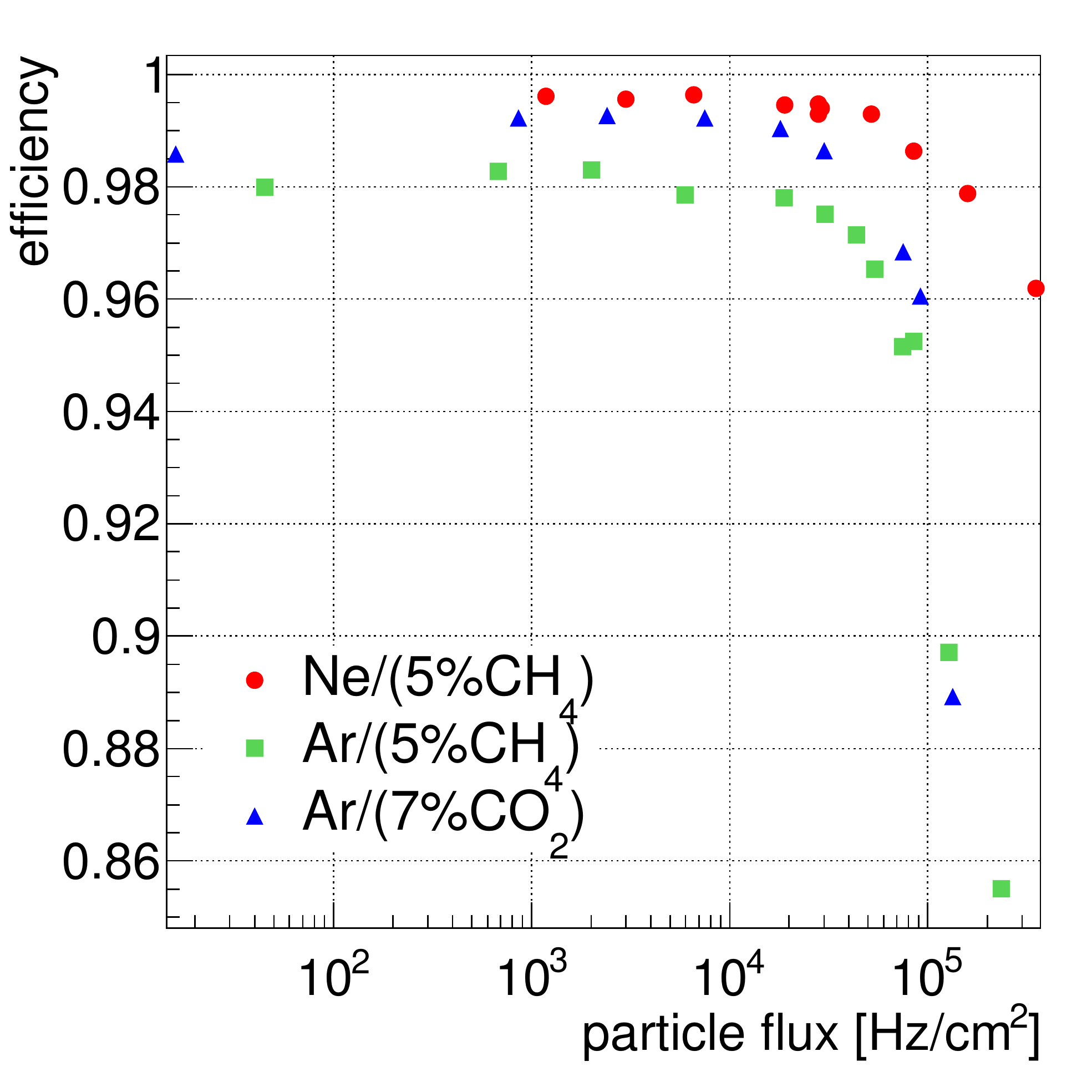}
\end{subfigure}\hspace{10mm}
\begin{subfigure}[t]{0.45\linewidth}\caption{}
\includegraphics[scale=0.35]{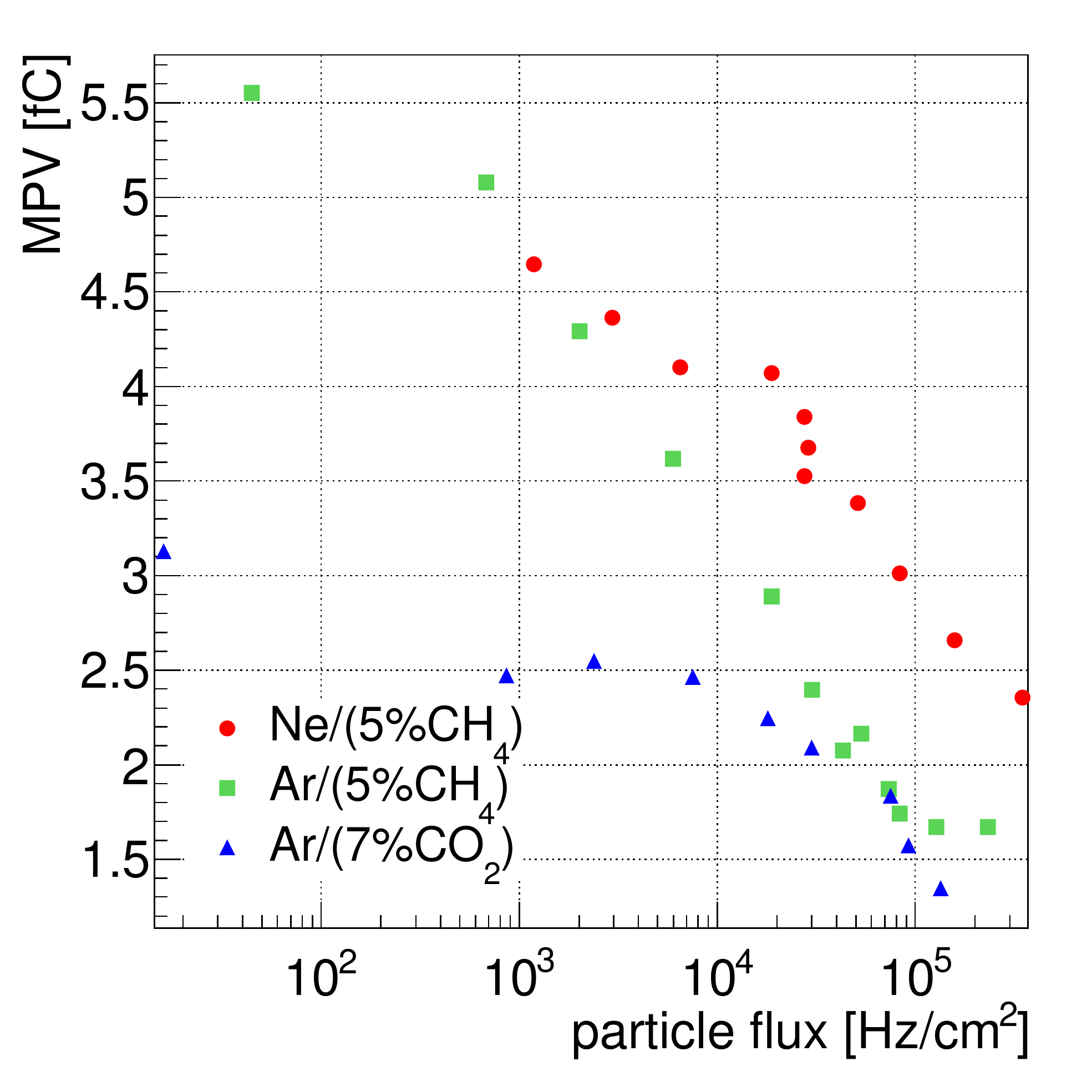}
\end{subfigure}
\caption{For the same data set: the global detection efficiency (a) and the Landau distribution MPV (b) as a function of the particle flux. \ldetsize RPWELL detector potential $\Delta$V$_{RPWELL}$ = 880~V, 1700~V, 1770~V in \nech~, \arch~, and \arco~ respectively.}\label{fig: Rate scan}
\end{figure}

\begin{figure}
\begin{subfigure}[b]{0.45\linewidth}\caption{}
\includegraphics[scale=0.35]{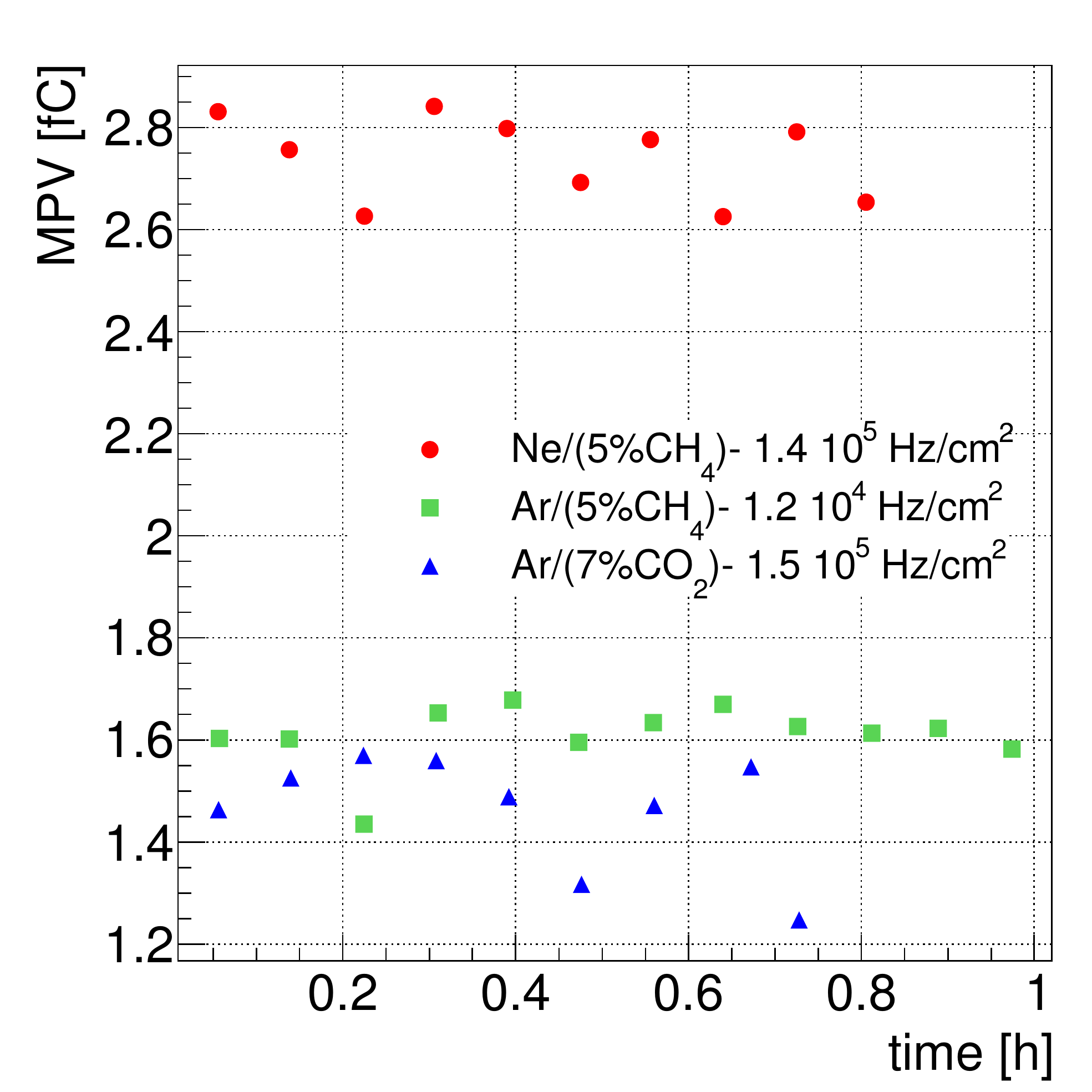}
\end{subfigure}\hspace{10mm}
\begin{subfigure}[b]{0.45\linewidth}\caption{}
\includegraphics[scale=0.35]{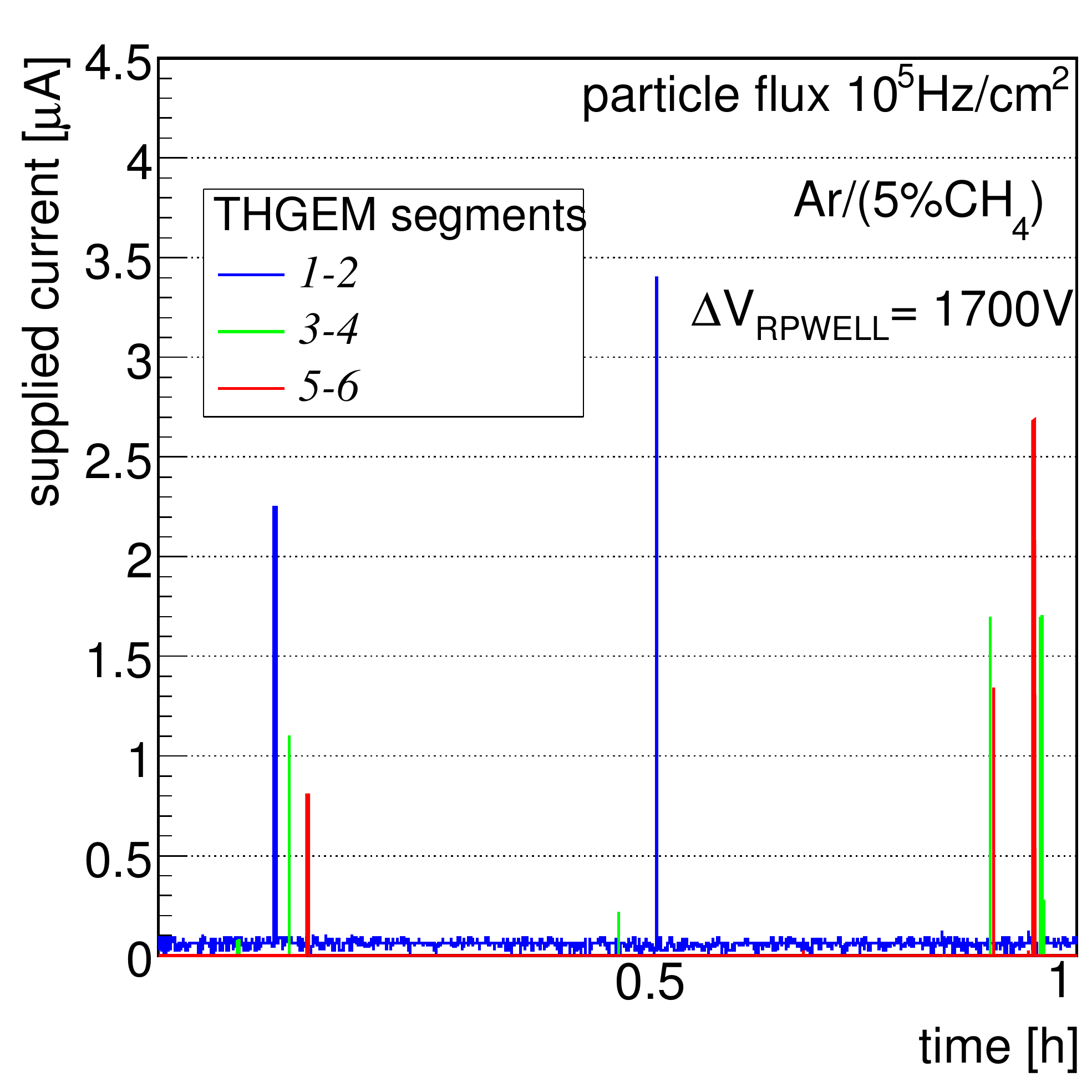}
\end{subfigure}
\caption{For the same data set: (a) Gain stability (MPV) as a function of time at high pion fluxes. \ldetsize RPWELL detector potential $\Delta$V$_{RPWELL}$ = 880~V, 1700~V, 1770~V in \nech~, \arch~, and \arco~ respectively. (b) Current supplied to the three couples of THGEM segments 1-2, 3-4, 5-6 (shown in figure~\ref{fig: assembly}-d), during 1 h operation at $\sim$10$^5$Hz/cm$^2$ particle flux in \arch. The beam was focused on the center of segment 3.}
\label{fig: Gain stability}
\end{figure}

The detector performance was investigated with low-rate muon and high rate pion beams; the latter reaching fluxes of $\sim$4$\cdot$10$^5$~Hz/cm$^2$. 
In order to maintain high detection efficiency at high particle fluxes, the measurements presented in this section were conducted under higher applied potential values compared to that optimized for the detection of low-rate muons (see table~\ref{tab: optimization}): $\Delta$V$_{RPWELL}$ was set to 880~V, 1700~V and 1770~V in \nech, \arch~and \arco~respectively. For all gas mixtures investigated, the global detection efficiency was stable up to rates of $\sim$10$^4$Hz/cm$^2$ (figure~\ref{fig: Rate scan}-a). The few percent efficiency drops at rates of $\sim$10$^5$Hz/cm$^2$ are due to $\sim$30\% gain loss at this rate compared to the gain measured at low-rates (figure~\ref{fig: Rate scan}-b); the loss in pulse-height can be attributed to charging up of the insulator within the holes and avalanche build-up limitations on the resistive anode (see for example~\cite{affatigato2015measurements}). These results are in agreement with that previously shown in~\cite{moleri2016resistive} for the \sdetsize RPWELL detector. If necessary, the efficiency drop can be partially mitigated using higher operation potentials. Only part of the efficiency lost can be recovered in this way, because a higher detector gain causes more charge to flow through the resistive layer and to charge up the insulator. In recent tests with a similar detector we verified this limitation, and we are now trying to quantify the effect.
The stability of the detector gain was investigated in the three gas mixtures, at pion fluxes of 10$^4$-10$^5$~Hz/cm$^2$; the results, in terms of the MPV as a function of time, are shown in Figure~\ref{fig: Gain stability}-a. No significant gain variations were observed along $\sim$1 hour of operation in all gas mixtures investigated.

\begin{figure}
\includegraphics[scale=0.35]{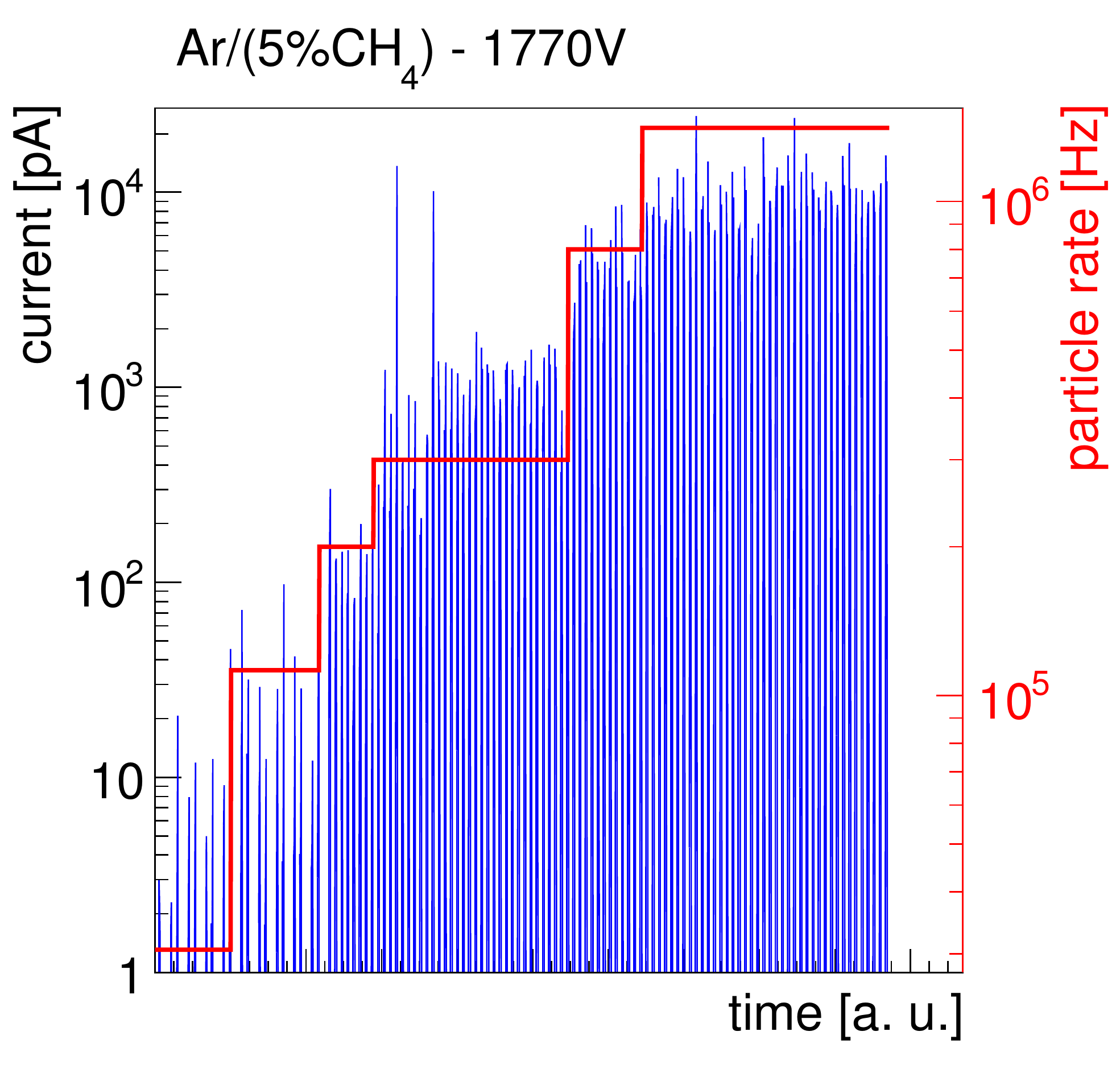}
\caption{Current flowing through the detector anode during pion runs at different rates in \arch. The beam spill structure is clearly visible.}\label{fig: discharges}
\end{figure} 

The same measurement was used to estimate the discharge probability. No discharges were observed 
when the detector was operated with \nech~and irradiated with >10$^8$ pions; therefore the resulting value of 10$^{-8}$ is an upper limit for the discharge probability in this gas mixture. In Argon based mixtures instead, under the considerably higher operation potentials, we observed sporadic discharges, as shown for example in figure~\ref{fig: Gain stability}-b, during the measurement in \arch. Note that such discharges were not observed under similar operation conditions in our \sdetsize detector prototype~\cite{moleri2016resistive}; also, since discharges were recorded also in segments located outside the beam area, we presume that they are most likely related to "weak points" in our modular detector prototype design: an open path along the support pins, leading to discharges propagating between the THGEM segment edge and the anode (see figure~\ref{fig: detector scheme}-c). This conclusion is supported by the sharp increase in the discharge probability measured when the rubber o-rings around each pin were absent (not presented here). This defect will be taken care of in future designs.

Figure~\ref{fig: discharges} shows the current flowing through the \ldetsize detector anode operated in \arco, under different pion rates, as a function of time; measured with a sensitive ammeter~\cite{Femtobox}. Similar results were obtained in all three gas mixtures. As expected, the small current spikes, corresponding to the beam spill structure, grow smoothly in amplitude with the particle rate. For all the rates investigated, the measured current is in agreement with the effective gain measured (figure~\ref{fig: Rate scan}-b);  $\mathrm{I= q\cdot n\cdot \Phi\cdot G(\Phi)}$, where I is the current, q is the electron charge, n is the number of electron-ion pairs produced by a minimum ionizing particle in 5~mm of Argon~\cite{sauli2014gaseous}, $\mathrm{\Phi}$ is the particle rate and G the detector gain (which depends on $\Phi$ as explained above).\footnote{As explained in~\cite{bressler2016first}, the total gain G($\Phi$) is about 5 times the measured effective gain.}
 
\section{Summary and discussion}
\label{sec: Summary and discussion}

A \ldetsize RPWELL detector prototype with a Semitron$^{\circledR}$ ESD225 resistive plate was assembled and tested. This thin, single-stage detector was operated with \nech~and with cost-effective \arch~and \arco~gas mixtures, at variable muon and pion fluxes. The operation in Argon mixtures would also permit having thinner drift gaps, possibly resulting in smaller inter-pad  multiplicities under inclined incidence. Its performance was compared with the one obtained with a \sdetsize prototype~\cite{moleri2016resistive}. Both prototypes demonstrated high detection efficiency (>98$\%$) at low average pad multiplicity ($\sim$1.2) in all three gas mixtures. The detection efficiency remained stable when the detector was exposed to particle fluxes up to 10$^4$Hz/cm$^2$ and dropped by few percent at 10$^5$Hz/cm$^2$. The current flowing through the detector anode increased with increasing particle flux and had no abnormal variations, indicating stable detector operation. The hexagonal THGEM hole pattern resulted in a somewhat higher local multiplicity at the inter-pad boundary, compared to that measured with a THGEM of a square hole-pattern~\cite{moleri2016resistive}. This, however, did not affect significantly the average pad multiplicity. This performance of the RPWELL detector is compatible with the requirements imposed for future digital hadron calorimetry; moreover, it is comparable or superior to that of other technologies suggested in this context (for example RPC, GEM, MICROMEGAS). A detailed comparison can be found in~\cite{moleri2016resistive}.
Discharge-free operation similar to the one shown in ~\cite{moleri2016resistive} for the \sdetsize prototype was demonstrated in \nech~gas mixture; occasional discharges were however observed in \arch~and \arco - associated with the support pins in the present design. Consequently, a different design is currently being implemented in a new \lldetsize detector prototype. The RPWELL concept is expected to pave ways towards various applications necessitating the deployment of robust large-area particle-imaging detectors.

\acknowledgments

This research was supported in part by the I-CORE Program of the Planning and Budgeting Committee, the Nella and Leon Benoziyo Center for High Energy Physics, a CERN-RD51 Common Project Grant and the European Union's Horizon 2020 research and innovation program under grant agreement No 654168. A. Breskin is the W.P. Reuther Professor of Research in the Peaceful use of Atomic Energy. F. D. Amaro acknowledges support by FCT under Post-Doctoral Grant SFRH/BPD/74775/2010. C. D. R. Azevedo acknowledges support by FCT under Post-Doctoral Grant SFRH/BPD//79163/2011.

\bibliographystyle{elsarticle-num}                 
\bibliography{bibliography.bib}

\end{document}